\documentclass[12pt]{article}

\usepackage{sbc-template}

\usepackage{graphicx,url}

\usepackage[utf8]{inputenc}

\usepackage{graphicx}
\usepackage{parcolumns}
\usepackage{color}
\usepackage{amsfonts}
\usepackage{colortbl}
\usepackage{booktabs}
\usepackage{amsmath}
\usepackage{tabularx}
\usepackage{textcomp}
\usepackage{soul}
\usepackage{footmisc}
\usepackage{changepage}
\usepackage[dvipsnames]{xcolor}
\usepackage[all]{nowidow}
\usepackage{enumerate}
\usepackage{multicol}
\usepackage{multirow}
\usepackage{authblk}
\usepackage{hyperref}
\usepackage{amsfonts}
\usepackage{subfigure}

\title{Microservices in Practice: A Survey Study}

\author{Markos Viggiato\inst{1}, Ricardo Terra\inst{2}, Henrique Rocha\inst{3}, Marco Tulio Valente\inst{1}, Eduardo Figueiredo\inst{1} }

\address{Dept. of Computer Science, Federal University of Minas Gerais (UFMG)\\
  Belo Horizonte -- MG -- Brazil
\nextinstitute
  Dept. of Computer Science, Federal University of Lavras (UFLA)\\
  Lavras -- MG -- Brazil
\nextinstitute
  Inria Lille - Nord Europe\\
  Villeneuve D'ascq, France
  \email{\{markosviggiato, mtov, figueiredo\}@dcc.ufmg.br, terra@dcc.ufla.br,
  henrique.rocha@inria.fr}
}

\begin{document} 

\maketitle

\begin{abstract}
Microservices architectures have become largely popular in the last years. However, we still lack empirical evidence about the use of microservices and the practices followed by practitioners. Thereupon, in this paper, we report the results of a survey with 122 professionals who work with microservices. We report how the industry is using this architectural style and whether the perception of practitioners regarding the advantages and challenges of microservices is according to the literature.

\end{abstract}

\section{Introduction}
\label{sec:intro}

Microservices have become largely popular in the last years together with the spread of \mbox{DevOps} practices and containers technologies, such as Kubernetes and Docker~\cite{pahl}. \textcolor{black}{We can see a significant increase in the use of microservices architectural style since 2014~\cite{workload}, which can be verified in the service-oriented software industry where the usage of microservices has been far superior when compared to other software architecture models~\cite{mappingStudy1-2016}.} 

Microservices are autonomous components that isolate fine-grained business capabilities. Furthermore, a microservice usually runs on its own process and communicates using standardized interfaces and lightweight protocols~\cite{jamesFowler,MicroAmbients}. In practice, microservices are widely used by large Web companies, such as Netflix, LinkedIn, and Amazon, which can be motivated by the benefits that microservices bring, e.g., the reduced time to put a new feature in operation~\cite{mappingStudy1-2016}.




\textcolor{black}{There are many benefits of using microservices, such as technology diversity in a single system, better scalability, increase productivity, and ease of deployment~\cite{mappingStudy1-2016,Newman:2015}. Consequently, these benefits may improve software maintainability~\cite{mappingStudy1-2016}. However, microservices also have their drawbacks. Usually, the services that compose the software are part of a distributed setting. Therefore, microservices could complicate some tasks such as finding a service within the network, managing the security, executing transactions, and optimizing the communication between services~\cite{mappingStudy1-2016,entreprise}.} 

Shedding light on microservices usage in practice is important for many reasons. 
We can perceive the advantages that motivate practitioners and the most important challenges faced when developing software under this architecture. This information can support decision-making about migrating systems to microservices or even start to develop an entire application under this architectural style. It can also aid software developers to understand and follow the best practices, making the microservices usage more effective. In addition, it may support the adoption of practices in software domains by practitioners and the developers' perception of the software quality~\cite{mori2018evaluating,oliveiraempirical,guimaraes2013prioritizing}. 

Nevertheless, there are no studies that investigate how microservices are used in practice. To the best of our knowledge, existing works consist of systematic mapping studies, which summarize the progress of microservices technology so far \cite{mappingStudy1-2016,mapping2Pooyan}. By contrast, in this paper, we propose to look at microservices from a practical perspective, i.e., with the aim of understanding and reveal how practitioners are in fact using microservices.
%
%
More specifically, we describe the results of a survey designed to reveal the usage of microservices in practice. First, we conducted a mapping study to identify and highlight potential advantages and challenges faced by professionals who work with microservices. Next, we surveyed developers about the findings of the mapping study, aiming at verifying whether the use of microservices in the industry is according to the best recommendations mentioned in the literature and whether the advantages and challenges found in the mapping study are in fact what practitioners face.




\section{Microservices}
\label{sec:microservices}


\textcolor{black}{Microservices are an architectural style in which the process of software development is done by using autonomous components that isolate fine-grained business functionalities and communicate one with other through standardized interfaces~\cite{MicroAmbients}. Due to an extensive use in web and cloud-based applications, we can observe a migration of some companies from the monolith architecture to microservices since the latter brings many benefits such as self-manageable (decentralized governance) and lightweight components~\cite{benchmarkingPooyan}.}

\textcolor{black}{The purpose of microservices is to use autonomous units that are isolated one from another and coordinate them into a distributed infrastructure by a lightweight container technology, such as Docker. Usually, the adoption of this architectural model implies also in adopting agile practice, such as \mbox{DevOps}, which reduces the time between implementing a change in the system and transferring this change to the production environment~\cite{benchmarkingPooyan}.}

\textcolor{black}{The isolation of business functionalities is highly recommended when using microservices, and allows independent development and deployment of each microservice. Moreover, the isolation also optimizes the autonomy and the replaceability of the services. Indeed, a microservice architectural style brings many benefits for developers but is also comes with many challenges. In Section~\ref{mapping}, we present a detailed description of the advantages and challenges of working with microservices.
}

\section{Study Design}
\label{sec:study}

This study included two phases, a mapping study (Section~\ref{mapping}), and a survey (Section~\ref{survey}), as described next. 

\subsection{Mapping Study}
\label{mapping}

Initially, we performed a mapping study to collect information about microservices from blogs and articles, as well as from more traditional literature, including books and papers. Mapping studies are particularly recommended for understanding emerging fields or technologies~\cite{Wohlin2012}, which is certainly the case of microservices. 
In order to retrieve documents about microservices, we used three specific search strings on Google: \textit{microservices architecture}, \textit{good features of microservices architecture}, and \textit{bad features/parts of microservices architecture}. Our intention was to gather documents regarding all aspects of microservices, such as documents proposing general terms and definitions (using the first string), and documents with the best characteristics and the main challenges faced by developers (using the second and third strings). After analyzing the retrieved documents, we identified five papers and one book from the scientific literature. Furthermore, we also considered 15 relevant documents from well-known experienced practitioners, including ten articles from websites and five documents from blogs. It is important to note that microservices have become popular in recent years, therefore there is still not a large number of relevant documents~available.

The first author of this paper carefully read all the 21 relevant documents in order to extract their recurrent topics and themes.
Thereupon, we classified the topics into advantages and challenges faced by developers when already using microservices architectures.





\underline{\textit{Advantages.}}
\textcolor{black}{In a system composed of multiple microservices, developers have the possibility of using many different technologies~\cite{Newman:2015}. This \textbf{technology diversity} is a very common characteristic in applications using microservices and it allows the use of the right tool for the right job. Moreover, the heterogeneity of technologies allows the addition of new technologies during development or maintenance in a more efficient way. For example, this is suitable for web applications where we can observe a constant and fast change in development environments and frameworks~\cite{marcoAngularJS}.} 
Another benefit often associated to microservices is the possibility of deploying a given service independently from the others. This \textbf{independent deployment} could lead to a faster implementation of new features~\cite{Newman:2015}. \textbf{Scalability} is also mentioned as an advantage of microservices since it can be achieved on demand, scaling only the service that contains a given functionality. It is also possible to replicate specific services, instead of the entire system. Finally, \textbf{maintainability} is often reported as a benefit since developers can modify or replace a service without impacting the entire application.



\underline{\textit{Challenges.}}
It is often reported that microservices demand distributed data management and hence \textbf{distributed transactions}, which makes their implementation much more complex. Other studies~\cite{mapping2Pooyan,benchmarkingPooyan} report that automated tests are extremely important in microservices, especially \textbf{integration tests}, which can be more complex and time-consuming. In addition, when the tests fail, it can be harder to determine which functionality has been broken~\cite{Newman:2015}. \textbf{Service faults} are also cited as a challenge when using microservices since the identification of a fault in a distributed setting is much harder than in a monolithic one. Finally, developers often mention that \textbf{Remote Procedure Calls} (RPC) are expensive and take much longer than local calls, which means that RPC may become a challenge when developing applications under microservices. 


\subsection{Survey Design}
\label{survey}

We designed a survey aiming at confirming (or not) the general characteristics, advantages, and challenges associated with microservices, as indicated by our mapping study. 
The survey has 14 questions and it is divided into three sections. The first section is about the background of the participants, including questions about experience with software development and with microservices as well as about the size of applications/number of services that they already worked with. The second section is related to definitions and trade-offs of using microservices. For instance, it includes questions about the ideal size of a microservice, and  advantages and problems faced by developers when using this technology. Finally, the third section is composed of open-ended questions about the technologies used by developers when implementing microservices applications. Our intention is to identify the most popular programming languages and technologies used under microservices architecture since the literature states that many technologies can be used in microservices systems.  


To find participants, we implemented an algorithm to search for microservices developers in the Stack Overflow community. We identified and retrieved nicknames from users who own questions or answers containing tag \textit{microservices}. In order to collect the email, we matched the Stack Overflow nickname with the equivalent nickname at GitHub. In addition, we promoted the survey in many communities about microservices and cloud-based development, including a Google group\footnote{https://groups.google.com/forum/\#!forum/microservices}, a Google Plus community\footnote{https://plus.google.com/communities/112442985624053749478}, and a Reddit community\footnote{https://www.reddit.com/r/microservices/}. The survey remained open between June and July 2017, and we obtained 122 complete responses.


\section{Survey Results}
\label{sec:results}

In this section, we describe the results obtained from the survey by presenting the participants background experience (Section~\ref{background}), the popular languages and technologies (Section~\ref{diversity}), the perceived advantages and challenges (Section~\ref{benefit}), and the participants' feedback (Section~\ref{sec:feed}).  

\subsection{Participants Background}
\label{background}

\begin{figure}[htbp]
\centering
\subfigure[Development experience]{\includegraphics[height=5.3cm]{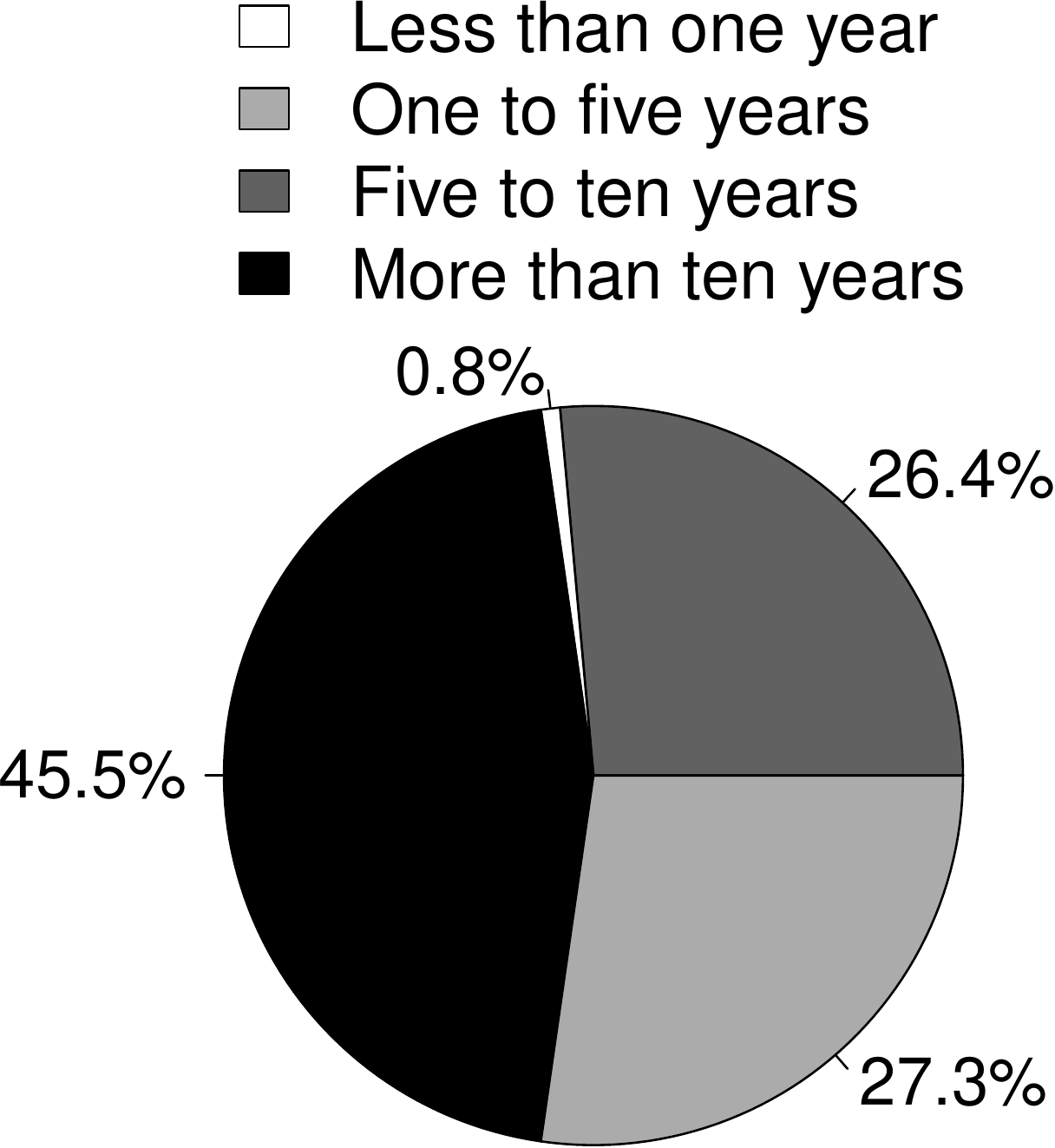}}\hspace*{30pt}
\subfigure[Microservices experience]{\includegraphics[height=5.3cm]{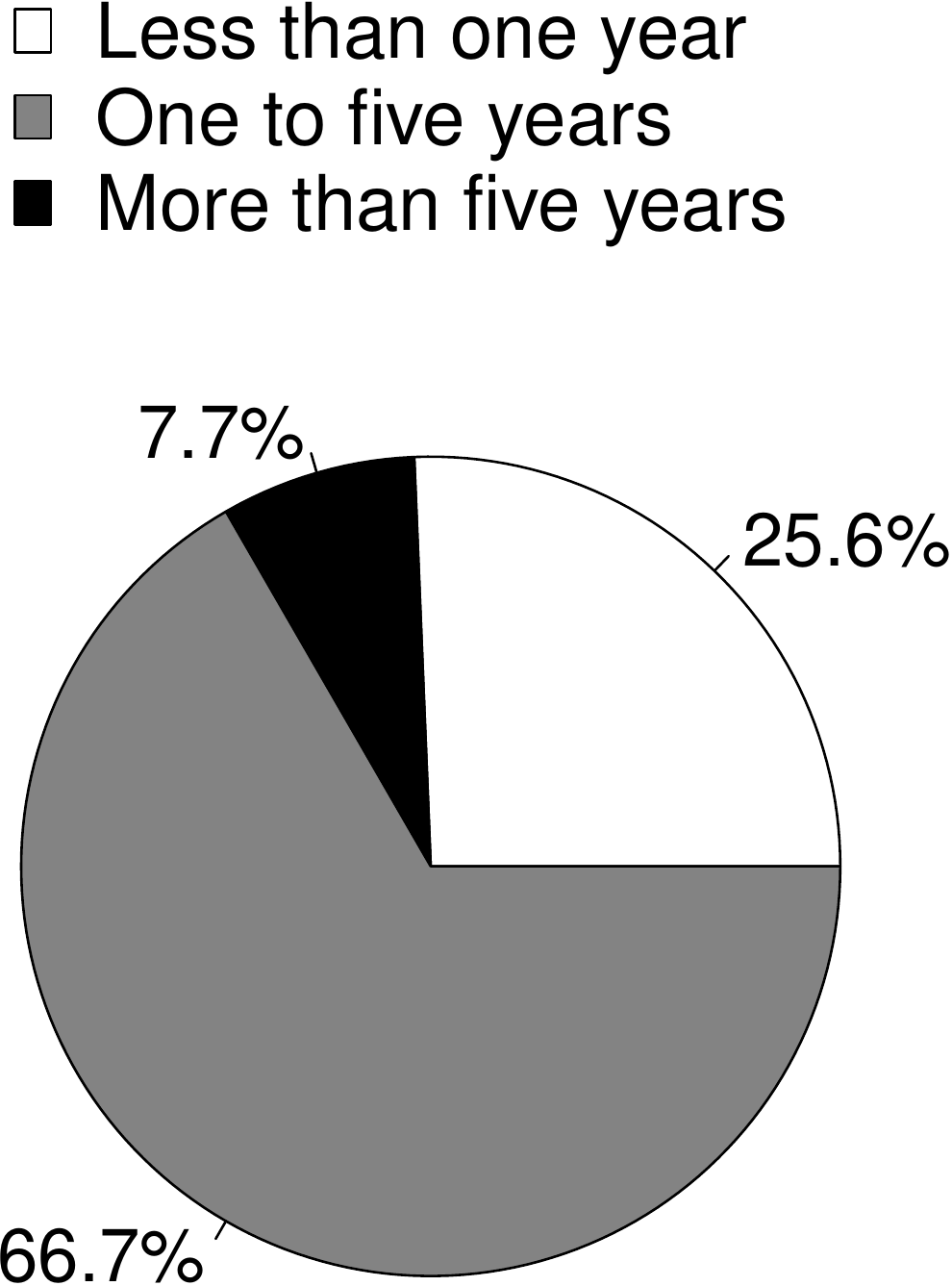}}\hspace*{30pt}
\subfigure[Participant's role]{\includegraphics[height=5.3cm]{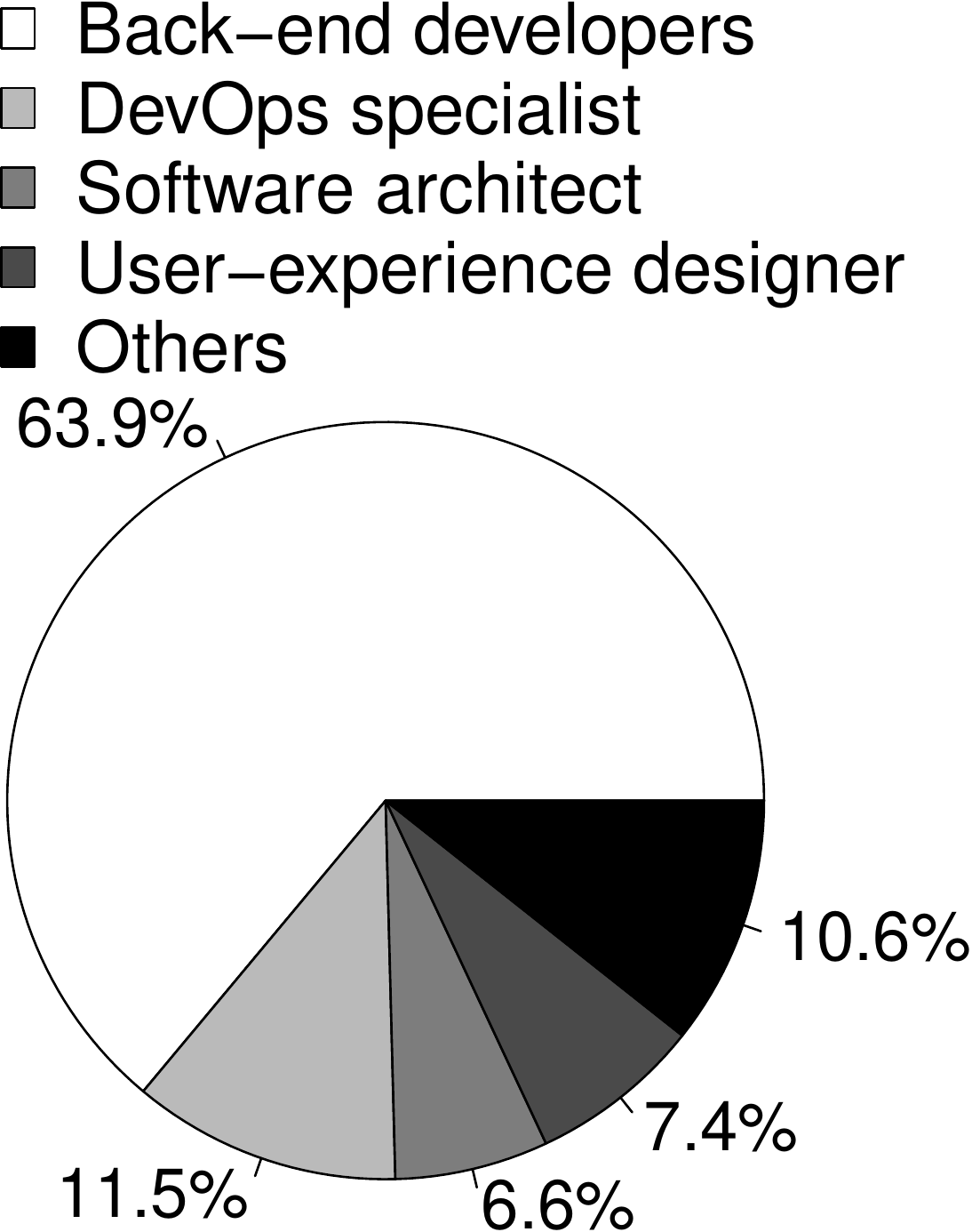}}\vspace*{10pt}
\caption{Participants' background. }
\label{fig:background}
\end{figure}

As presented in Figure~\ref{fig:background}, almost 72\% of the participants have at least five years of experience with software development. Furthermore, about 74\% of them have more than one year of experience with microservices; more specifically, almost 67\% have one to five years of experience. In addition, approximately 64\% of the respondents are back-end developers while only 11.5\% are DevOps. 

Although not graphically illustrated, we also collected data regarding the size of the applications. 70.5\% of the respondents worked with monolithic-based systems larger than 50~KLOC, and about 54\% worked with microservices-based applications larger than 10~KLOC. These numbers confirm that most survey participants are not novice in the microservices field.

Considering that microservices are an emerging field, and based on the professional background of the respondents, we claim the participants have sufficient knowledge on the subject to answer the survey questions.

\subsection{Most Popular Programming Languages and Technologies}
\label{diversity}

Aiming at characterizing microservices applications, we asked the survey participants about the languages and technologies they usually use in their projects. We found that four programming languages are largely used: Java (33\%), JavaScript through Node.js (18\%), C\#~(12\%), and PHP (8\%). The answers also mention other 14 programming languages, which indicate the flexibility of microservices-based applications regarding programming languages.
Regarding  the most common DBMS, the results include Postgres (30\%), MySQL (25\%), MongoDB (20\%), SQL Server (12\%), and Oracle (9\%). Finally, regarding the communication protocols, 62\% of the participants declared they use REST over HTTP. 

\subsection{Advantages and Challenges of Microservices}
\label{benefit}

\begin{figure}[h!]
\begin{minipage}{\textwidth}

\centering
\subfigure[Advantages]{\includegraphics[width=0.990\textwidth]{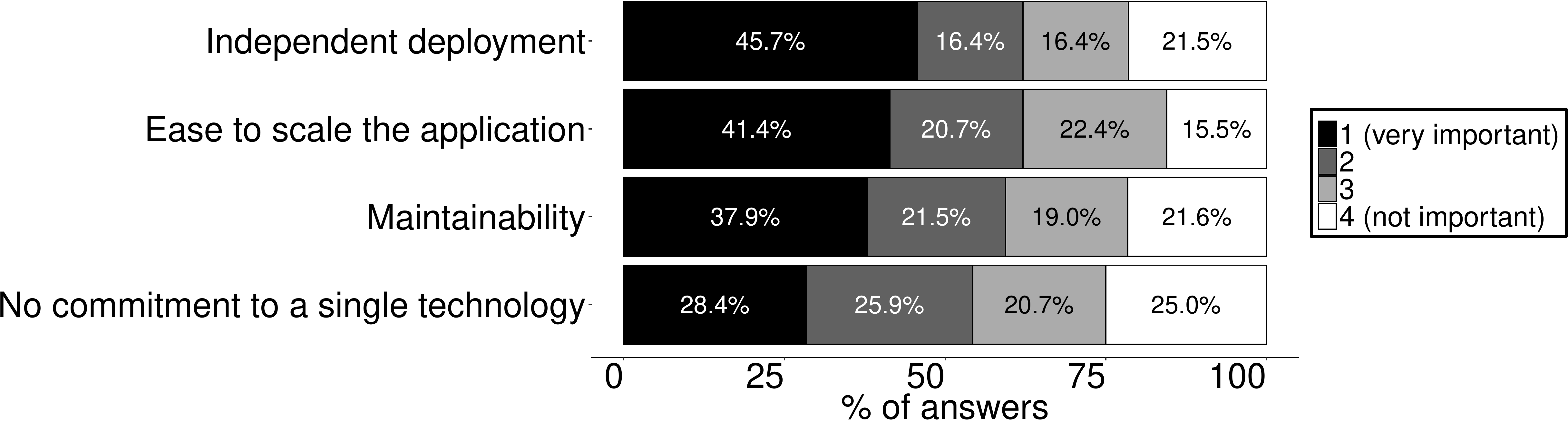}}\\[1cm]
\subfigure[Challenges]{\includegraphics[width=0.990\textwidth]{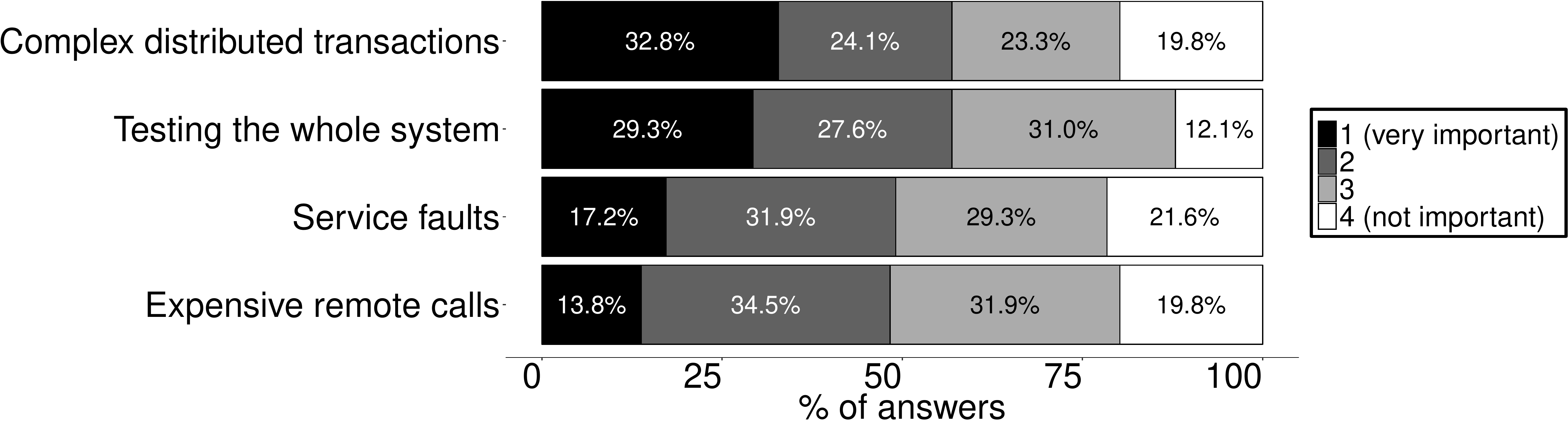}}
\vspace{10pt}
\caption{Microservices advantages and challenges}
\vspace{10pt}
\label{fig:adv_probl}

\end{minipage}
\end{figure}

Figure~\ref{fig:adv_probl}a presents the percentage of participants' answers about four advantages usually associated to microservices, in a scale from 1 (very important) to 4 (not important at all). For \textbf{independent deployment}, we can see the largest difference from score 1 to the others, which reveals a higher agreement rate for this feature when compared to the others. Yet, for the other three advantages, we can verify that more than 50\% of the respondents chose scores~1 or 2, indicating these characteristics are in fact relevant when working~\mbox{with~microservices}.

From the mapping study, we also identified four main challenges faced by developers when working with microservices. In Figure~\ref{fig:adv_probl}b, we can see the responses of the survey's participants regarding these challenges. The participants agree that \textbf{complex distributed transactions} are a very important challenge. In fact, this challenge has the highest percentage for score~1~(32.8\%). We can also observe that about 57\% and 49\% consider \textbf{testing the whole system} and \textbf{service faults}, respectively, as important challenges (scores 1 and~2). In contrast, challenge \textbf{expensive remote calls} is very important to only 13.8\% of the participants. Interestingly, practitioners disagree with the literature in this last challenge since they do not find expensive remote calls a very important challenge in microservices development. In a nutshell, developers should pay special attention to distributed transactions, testing of the whole system and service faults, since these may become big problems in the system.


\subsection{Feedback from Participants}
\label{sec:feed}

We also sent another email to the developers directly contacted to answer our survey. In this follow-up message, we described the major survey results with the aim of receiving their impressions about our study and findings. We also intended to verify whether they agree or not with our results. We received nine answers; all of them with a positive feedback.
In general, developers answered that our paper provides a good overview of the microservice practice. For instance, two developers commented this is {\em really interesting} and {\em good paper}. Another developer highlighted that microservices are not a ``holy grail'' and that monolithic can also have small and separated modules, even in distributed settings. 

\section{Threats to Validity}
\label{sec:threats}

Some threats may affect the validity of our findings. First, the survey parti\-ci\-pants may not represent the entire population of microservices practitioners. To mitigate this risk, we put efforts in promoting the survey in many different communities to include professionals from different software ecosystems. Second, the term {\em DevOps} (which is one of the options of the survey question about the participant's role) might not be common in some contexts. For example, in small organizations, DevOps tasks such as development and delivery process automation may fall on senior developers and architects. 
Third, although it would be desirable for our survey the analysis of larger (w.r.t.~size), stable (w.r.t.~age), and specific branches of industry applications, we argue that our survey brings a broad overview since it is based on the expertise of 122 developers who work with heterogeneous microservices-based applications.

\section{Related Work}
\label{sec:related}


\textcolor{black}{Most of the research in microservices restrict their study to a specific domain, such as business~\cite{entreprise}. There are also researches investigating microservices by performing a systematic mapping study. For instance, a study summarized the progress of studies about microservices until 2016, and identified the gaps and requirements~\cite{mappingStudy1-2016}. Other study taxonomically classified and compared studies of this architectural style and their application in the cloud~\cite{mapping2Pooyan}. Our study follows a different route as we look at the microservices from a practical perspective. We aimed to understand and indicate how the software development industry is, in fact, using this popular architecture, and how practitioners perceive the advantages and challenges of microservices. 
} 


\section{Conclusion}
\label{sec:conclusion}

The findings of this paper indicate that 
practitioners usually follow 
the best practices for microservices reported in the literature.
We also confirmed the benefits provided by microservices, such as \textbf{independent deployment}, \textbf{ease to scale the applications}, \textbf{maintainability}, and \textbf{no commitment to a single technology stack}.
Last but not least, we also confirmed the challenges developers may face, such as \textbf{complex distributed transactions}, \textbf{testing the whole system}, and \textbf{service faults}.

However, we also found some important topics that are in disagreement. First, professionals usually work as back-end developers (64\%), instead of as \textbf{DevOps} specialists in cross-functional teams.
Second, according to the mapping study, \textbf{expensive remote calls} is one of the challenges that developers face when working with microservices. However, about 52\% of the respondents declared that it is not an important or it is a little important issue in their systems.

As future work, we intend to conduct interviews with microservices professionals to confirm our results and to better understand if and why practitioners do not follow some best practices. We also plan to perform an industrial-scale case study with companies that adopt microservices to monitor real software developers developing microservices-based projects in order to report the problems they face and the solutions they apply.


\section* {Acknowledgements} 
Our research has been supported by CAPES, FAPEMIG, and CNPq.


\bibliographystyle{sbc}
\bibliography{9-references}

\end{document}